# A CORRELATED TOPIC MODEL OF SCIENCE[1]

By David M. Blei and John D. Lafferty

*Princeton University and Carnegie Mellon University*

Topic models, such as latent Dirichlet allocation (LDA), can be useful tools for the statistical analysis of document collections and other discrete data. The LDA model assumes that the words of each document arise from a mixture of *topics*, each of which is a distribution over the vocabulary. A limitation of LDA is the inability to model topic correlation even though, for example, a document about genetics is more likely to also be about disease than X-ray astronomy. This limitation stems from the use of the Dirichlet distribution to model the variability among the topic proportions. In this paper we develop the correlated topic model (CTM), where the topic proportions exhibit correlation via the logistic normal distribution [*J. Roy. Statist. Soc. Ser. B* **44** (1982) 139–177]. We derive a fast variational inference algorithm for approximate posterior inference in this model, which is complicated by the fact that the logistic normal is not conjugate to the multinomial. We apply the CTM to the articles from *Science* published from 1990–1999, a data set that comprises 57M words. The CTM gives a better fit of the data than LDA, and we demonstrate its use as an exploratory tool of large document collections.

## 1. Introduction.

Large collections of documents are readily available on-line and widely accessed by diverse communities. As a notable example, scholarly articles are increasingly published in electronic form, and historical archives are being scanned and made accessible. The not-for-profit organization JSTOR (www.jstor.org) is currently one of the leading providers of journals to the scholarly community. These archives are created by scanning old journals and running an optical character recognizer over the pages. JSTOR provides the original scans on-line, and uses their noisy version of

Received March 2007; revised April 2007.

[1]Supported in part by NSF Grants IIS-0312814 and IIS-0427206, the DARPA CALO project and a grant from Google.

Supplementary material and code are available at http://imstat.org/aoas/supplements

the text to support keyword search. Since the data are largely unstructured and comprise millions of articles spanning centuries of scholarly work, automated analysis is essential. The development of new tools for browsing, searching and allowing the productive use of such archives is thus an important technological challenge, and provides new opportunities for statistical modeling.

The statistical analysis of documents has a tradition that goes back at least to the analysis of the Federalist papers by Mosteller and Wallace [21]. But document modeling takes on new dimensions with massive multi-author collections such as the large archives that now are being made accessible by JSTOR, Google and other enterprises. In this paper we consider *topic models* of such collections, by which we mean latent variable models of documents that exploit the correlations among the words and latent semantic themes. Topic models can extract surprisingly interpretable and useful structure without any explicit "understanding" of the language by computer. In this paper we present the *correlated topic model* (CTM), which explicitly models the correlation between the latent topics in the collection, and enables the construction of topic graphs and document browsers that allow a user to navigate the collection in a topic-guided manner.

The main application of this model that we present is an analysis of the JSTOR archive for the journal *Science*. This journal was founded in 1880 by Thomas Edison and continues as one of the most influential scientific journals today. The variety of material in the journal, as well as the large number of articles ranging over more than 100 years, demonstrates the need for automated methods, and the potential for statistical topic models to provide an aid for navigating the collection.

The correlated topic model builds on the earlier latent Dirichlet allocation (LDA) model of Blei, Ng and Jordan [8], which is an instance of a general family of mixed membership models for decomposing data into multiple latent components. LDA assumes that the words of each document arise from a mixture of topics, where each topic is a multinomial over a fixed word vocabulary. The topics are shared by all documents in the collection, but the topic proportions vary stochastically across documents, as they are randomly drawn from a Dirichlet distribution. Recent work has used LDA as a building block in more sophisticated topic models, such as author-document models [19, 24], abstract-reference models [12] syntax-semantics models [16] and image-caption models [6]. The same kind of modeling tools have also been used in a variety of nontext settings, such as image processing [13, 26], recommendation systems [18], the modeling of user profiles [14] and the modeling of network data [1]. Similar models were independently developed for disability survey data [10, 11] and population genetics [22].

In the parlance of the information retrieval literature, LDA is a "bag of words" model. This means that the words of the documents are assumed to be exchangeable within them, and Blei, Ng and Jordan [8] motivate LDA from this assumption with de Finetti's exchangeability theorem. As a consequence, models like LDA represent documents as vectors of word counts in a very high dimensional space, ignoring the order in which the words appear. While it is important to retain the exact sequence of words for reading comprehension, the linguistically simplistic exchangeability assumption is essential to efficient algorithms for automatically eliciting the broad semantic themes in a collection.

The starting point for our analysis here is a perceived limitation of topic models such as LDA: they fail to directly model correlation between topics. In most text corpora, it is natural to expect that subsets of the underlying latent topics will be highly correlated. In *Science*, for instance, an article about genetics may be likely to also be about health and disease, but unlikely to also be about X-ray astronomy. For the LDA model, this limitation stems from the independence assumptions implicit in the Dirichlet distribution on the topic proportions. Under a Dirichlet, the components of the proportions vector are nearly independent, which leads to the strong and unrealistic modeling assumption that the presence of one topic is not correlated with the presence of another. The CTM replaces the Dirichlet by the more flexible logistic normal distribution, which incorporates a covariance structure among the components [4]. This gives a more realistic model of the latent topic structure where the presence of one latent topic may be correlated with the presence of another.

However, the ability to model correlation between topics sacrifices some of the computational conveniences that LDA affords. Specifically, the conjugacy between the multinomial and Dirichlet facilitates straightforward approximate posterior inference in LDA. That conjugacy is lost when the Dirichlet is replaced with a logistic normal. Standard simulation techniques such as Gibbs sampling are no longer possible, and Metropolis–Hastings based MCMC sampling is prohibitive due to the scale and high dimension of the data.

Thus, we develop a fast variational inference procedure for carrying out approximate inference with the CTM model. Variational inference [17, 29] trades the unbiased estimates of MCMC procedures for potentially biased but computationally efficient algorithms whose numerical convergence is easy to assess. Variational inference algorithms have been effective in LDA for analyzing large document collections [8]. We extend their use to the CTM.

The rest of this paper is organized as follows. We first present the correlated topic model and discuss its underlying modeling assumptions. Then, we present an outline of the variational approach to inference (the technical details are collected in the Appendix) and the variational expectation–maximization procedure for parameter estimation. Finally, we analyze the performance of the model on the JSTOR *Science* data. Quantitatively, we show that it gives a better fit than LDA, as measured by the accuracy of the predictive distributions over held out documents. Qualitatively, we present an analysis of all of *Science* from 1990–1999, including examples of topically related articles found using the inferred latent structure, and topic graphs that are constructed from a sparse estimate of the covariance structure of the model. The paper concludes with a brief discussion of the results and future work that it suggests.

**2. The correlated topic model.** The *correlated topic model* (CTM) is a hierarchical model of document collections. The CTM models the words of each document from a mixture model. The mixture components are shared by all documents in the collection; the mixture proportions are document-specific random variables. The CTM allows each document to exhibit multiple topics with different proportions. It can thus capture the heterogeneity in grouped data that exhibit multiple latent patterns.

We use the following terminology and notation to describe the data, latent variables and parameters in the CTM.

- **Words and documents.** The only observable random variables that we consider are *words* that are organized into *documents*. Let $w_{d,n}$ denote the $n$th word in the $d$th document, which is an element in a $V$-term vocabulary. Let $\boldsymbol{w}_d$ denote the vector of $N_d$ words associated with document $d$.
- **Topics.** A *topic* $\boldsymbol{\beta}$ is a distribution over the vocabulary, a point on the $V-1$ simplex. The model will contain $K$ topics $\boldsymbol{\beta}_{1:K}$.
- **Topic assignments.** Each word is each assumed drawn from one of the $K$ topics. The *topic assignment* $z_{d,n}$ is associated with the $n$th word and $d$th document.
- **Topic proportions.** Finally, each document is associated with a set of *topic proportions* $\boldsymbol{\theta}_d$, which is a point on the $K-1$ simplex. Thus, $\boldsymbol{\theta}_d$ is a distribution over topic indices, and reflects the probabilities with which words are drawn from each topic in the collection. We will typically consider a natural parameterization of this multinomial $\boldsymbol{\eta} = \log(\theta_i/\theta_K)$.

Specifically, the correlated topic model assumes that an $N$-word document arises from the following generative process. Given topics $\boldsymbol{\beta}_{1:K}$, a $K$-vector $\boldsymbol{\mu}$ and a $K \times K$ covariance matrix $\boldsymbol{\Sigma}$:

1. Draw $\boldsymbol{\eta}_d | \{\boldsymbol{\mu}, \boldsymbol{\Sigma}\} \sim \mathcal{N}(\boldsymbol{\mu}, \boldsymbol{\Sigma})$.
2. For $n \in \{1, \dots, N_d\}$:

(a) Draw topic assignment $Z_{d,n} | \boldsymbol{\eta}_d$ from $\mathrm{Mult}(f(\boldsymbol{\eta}_d))$.
(b) Draw word $W_{d,n} | \{z_{d,n}, \boldsymbol{\beta}_{1:K}\}$ from $\mathrm{Mult}(\boldsymbol{\beta}_{z_{d,n}})$,

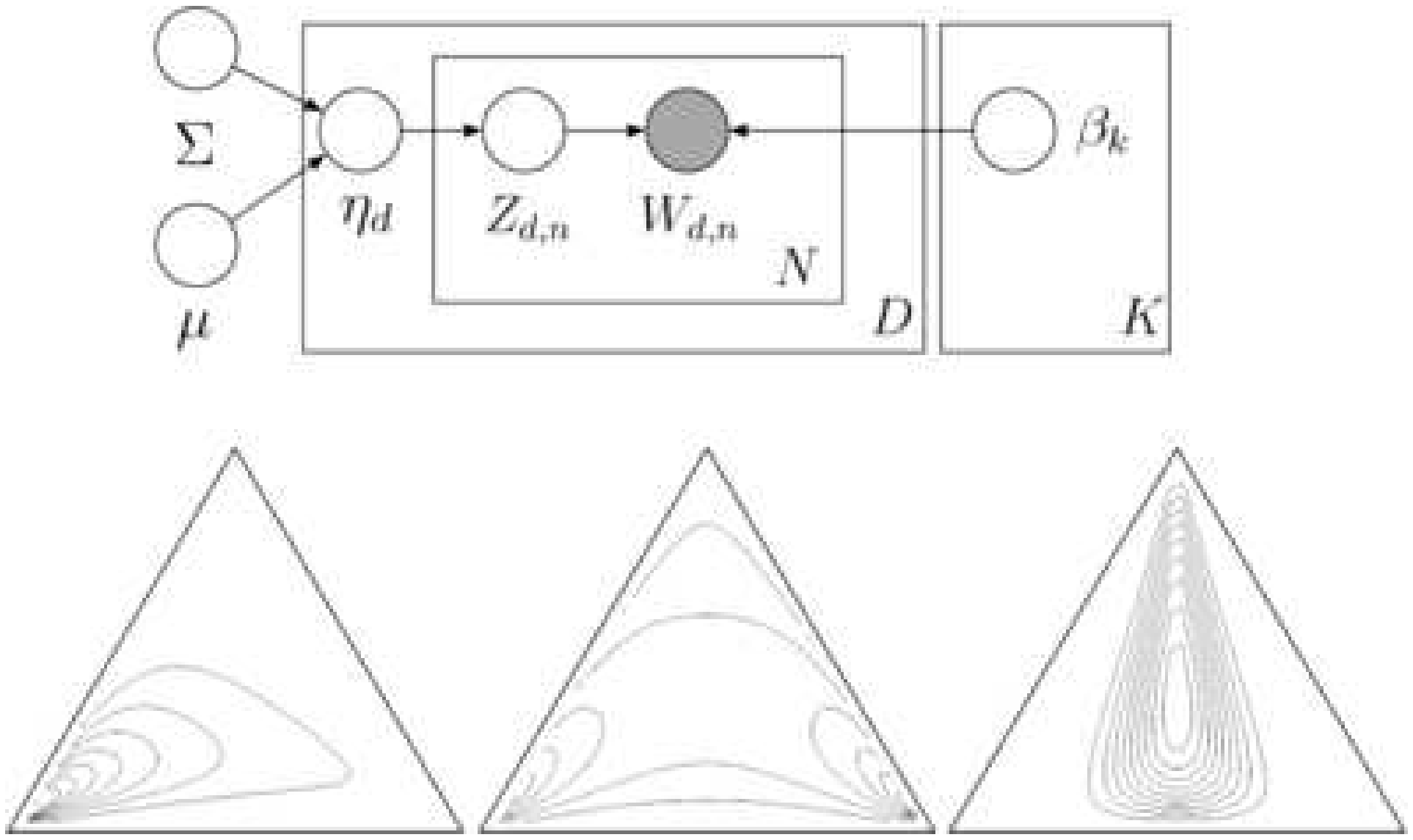

FIG. 1. Top: *Probabilistic graphical model representation of the correlated topic model. The logistic normal distribution, used to model the latent topic proportions of a document, can represent correlations between topics that are impossible to capture using a Dirichlet.* Bottom: *Example densities of the logistic normal on the* 2*-simplex. From left: diagonal covariance and nonzero-mean, negative correlation between topics* 1 *and* 2*, positive correlation between topics* 1 *and* 2.

where $f(\boldsymbol{\eta})$ maps a natural parameterization of the topic proportions to the mean parameterization,

$$\boldsymbol{\theta} = f(\boldsymbol{\eta}) = \frac{\exp\{\boldsymbol{\eta}\}}{\sum_i \exp\{\eta_i\}}. \tag{1}$$

(Note that $\boldsymbol{\eta}$ does not index a minimal exponential family. Adding any constant to $\boldsymbol{\eta}$ will result in an identical mean parameter.) This process is illustrated as a probabilistic graphical model in Figure 1. (A *probabilistic graphical model* is a graph representation of a family of joint distributions with a graph. Nodes denote random variables; edges denote possible dependencies between them.)

The CTM builds on the latent Dirichlet allocation (LDA) model [8]. Latent Dirichlet allocation assumes a nearly identical generative process, but one where the topic proportions are drawn from a Dirichlet. In LDA and its variants, the Dirichlet is a computationally convenient distribution over topic proportions because it is conjugate to the topic assignments. But, the Dirichlet assumes near independence of the components of the proportions. In fact, one can simulate a draw from a Dirichlet by drawing from $K$ independent Gamma distributions and normalizing the resulting vector. (Note that there is slight negative correlation due to the constraint that the components sum to one.)

Rather than use a Dirichlet, the CTM draws a real valued random vector from a multivariate Gaussian and then maps it to the simplex to obtain a multinomial parameter. This is the defining characteristic of the logistic Normal distribution [2, 3, 4]. The covariance of the Gaussian induces dependencies between the components of the transformed random simplicial vector, allowing for a general pattern of variability between its components. The logistic normal was originally studied in the context of analyzing observed compositional data, such as the proportions of minerals in geological samples. In the CTM, we use it to model the *latent* composition of topics associated with each document.

The drawback of using the logistic normal is that it is not conjugate to the multinomial, which complicates the corresponding approximate posterior inference procedure. The advantage, however, is that it provides a more expressive document model. The strong independence assumption imposed by the Dirichlet is not realistic when analyzing real document collections, where one finds strong correlations between the latent topics. For example, a document about geology is more likely to also be about archeology than genetics. We aim to use the covariance matrix of the logistic normal to capture such relationships.

In Section 4 we illustrate how the higher order structure given by the covariance can be used as an exploratory tool for better understanding and navigating a large corpus of documents. Moreover, modeling correlation can lead to better predictive distributions. In some applications, such as automatic recommendation systems, the goal is to predict unseen items conditioned on a set of observations. A Dirichlet-based model will predict items based on the latent topics that the observations suggest, but the CTM will predict items associated with *additional* topics that are correlated with the conditionally probable topics.

**3. Computation with the correlated topic model.** We address two computational problems that arise when using the correlated topic model to analyze data. First, given a collection of topics and distribution over topic proportions $\{\boldsymbol{\beta}_{1:K}, \boldsymbol{\mu}, \boldsymbol{\Sigma}\}$, we estimate the posterior distribution of the latent variables conditioned on the words of a document $p(\boldsymbol{\eta}, \boldsymbol{z} | \boldsymbol{w}, \boldsymbol{\beta}_{1:K}, \boldsymbol{\mu}, \boldsymbol{\Sigma})$. This lets us embed newly observed documents into the low dimensional latent thematic space that the model represents. We use a fast variational inference algorithm to approximate this posterior, which lets us quickly analyze large document collections under these complicated modeling assumptions.

Second, given a collection of documents $\{\boldsymbol{w}_1, \ldots, \boldsymbol{w}_D\}$, we find maximum likelihood estimates of the topics and the underlying logistic normal distribution under the modeling assumptions of the CTM. We use a variant of the expectation-maximization algorithm, where the E-step is the per-document posterior inference problem described above. Furthermore, we seek sparse solutions of the inverse covariance matrix between topics, and we adapt $\ell_1$-regularized covariance estimation [20] for this purpose.

3.1. *Posterior inference with variational methods.* Given a document $\boldsymbol{w}$ and a model $\{\boldsymbol{\beta}_{1:K}, \boldsymbol{\mu}, \boldsymbol{\Sigma}\}$, the posterior distribution of the per-document latent variables is

$$\begin{aligned} &p(\boldsymbol{\eta}, \boldsymbol{z} | \boldsymbol{w}, \boldsymbol{\beta}_{1:K}, \boldsymbol{\mu}, \boldsymbol{\Sigma}) \\ &\quad = \frac{p(\boldsymbol{\eta}|\boldsymbol{\mu}, \boldsymbol{\Sigma}) \prod_{n=1}^{N} p(z_n|\boldsymbol{\eta}) p(w_n|z_n, \boldsymbol{\beta}_{1:K})}{\int p(\boldsymbol{\eta}|\boldsymbol{\mu}, \boldsymbol{\Sigma}) \prod_{n=1}^{N} \sum_{z_n=1}^{K} p(z_n|\boldsymbol{\eta}) p(w_n|z_n, \boldsymbol{\beta}_{1:K})\, d\boldsymbol{\eta}}, \end{aligned} \tag{2}$$

which is intractable to compute due to the integral in the denominator, that is, the marginal probability of the document that we are conditioning on. There are two reasons for this intractability. First, the sum over the $K$ values of $z_n$ occurs inside the product over words, inducing a combinatorial number of terms. Second, even if $K^N$ stays within the realm of computational tractability, the distribution of topic proportions $p(\boldsymbol{\eta}|\boldsymbol{\mu}, \boldsymbol{\Sigma})$ is not conjugate to the distribution of topic assignments $p(z_n|\boldsymbol{\eta})$. Thus, we cannot analytically compute the integrals of each term.

The nonconjugacy further precludes using many of the Monte Carlo Markov chain (MCMC) sampling techniques that have been developed for computing with Dirichlet-based mixed membership models [10, 15]. These MCMC methods are all based on Gibbs sampling, where the conjugacy between the latent variables lets us compute coordinate-wise posteriors analytically. To employ MCMC in the logistic normal setting considered here, we have to appeal to a tailored Metropolis–Hastings solution. Such a technique will not enjoy the same convergence properties and speed of the Gibbs samplers, which is particularly hindering for the goal of analyzing collections that comprise millions of words.

Thus, to approximate this posterior, we appeal to variational methods as a deterministic alternative to MCMC. The idea behind variational methods is to optimize the free parameters of a distribution over the latent variables so that the distribution is close in Kullback–Leibler divergence to the true posterior [17, 29]. The fitted *variational distribution* is then used as a substitute for the posterior, just as the empirical distribution of samples is used in MCMC. Variational methods have had widespread application in machine learning; their potential in applied Bayesian statistics is beginning to be realized.

In models composed of conjugate-exponential family pairs and mixtures, the variational inference algorithm can be automatically derived by computing expectations of natural parameters in the variational distribution [5, 7, 31]. However, the nonconjugate pair of variables in the CTM requires that we derive the variational inference algorithm from first principles.

We begin by using Jensen's inequality to bound the log probability of a document,

$$
\begin{aligned}
&\log p(w_{1:N}|\boldsymbol{\mu}, \boldsymbol{\Sigma}, \boldsymbol{\beta}) \\
&\quad \geq \mathrm{E}_q[\log p(\boldsymbol{\eta}|\boldsymbol{\mu}, \boldsymbol{\Sigma})] + \sum_{n=1}^{N} \mathrm{E}_q[\log p(z_n|\boldsymbol{\eta})] \\
&\qquad + \sum_{n=1}^{N} \mathrm{E}_q[\log p(w_n|z_n, \boldsymbol{\beta})] + \mathrm{H}(q),
\end{aligned} \tag{3}
$$

where the expectation is taken with respect to $q$, a variational distribution of the latent variables, and $\mathrm{H}(q)$ denotes the entropy of that distribution. As a variational distribution, we use a fully factorized model, where all the variables are independently governed by a different distribution,

$$
q(\eta_{1:K}, z_{1:N}|\lambda_{1:K}, \nu^2_{1:K}, \boldsymbol{\phi}_{1:N}) = \prod_{i=1}^{K} q(\eta_i|\lambda_i, \nu_i^2) \prod_{n=1}^{N} q(z_n|\boldsymbol{\phi}_n). \tag{4}
$$

The variational distributions of the discrete topic assignments $z_{1:N}$ are specified by the $K$-dimensional multinomial parameters $\boldsymbol{\phi}_{1:N}$ (these are mean parameters of the multinomial). The variational distribution of the continuous variables $\eta_{1:K}$ are $K$ independent univariate Gaussians $\{\lambda_i, \nu_i\}$. Since the variational parameters are fit using a *single* observed document $w_{1:N}$, there is no advantage in introducing a nondiagonal variational covariance matrix.

The variational inference algorithm optimizes equation (3) with respect to the variational parameters, thereby tightening the bound on the marginal probability of the observations as much as the structure of variational distribution allows. This is equivalent to finding the variational distribution that minimizes $\mathrm{KL}(q||p)$, where $p$ is the true posterior [17, 29]. Details of this optimization for the CTM are given in the Appendix.

Note that variational methods do not come with the same theoretical guarantees as MCMC, where the limiting distribution of the chain is exactly the posterior of interest. However, variational methods provide fast algorithms and a clear convergence criterion, whereas MCMC methods can be computationally inefficient and determining when a Markov chain has converged is difficult [23].

3.2. *Parameter estimation.* Given a collection of documents, we carry out parameter estimation for the correlated topic model by attempting to maximize the likelihood of a corpus of documents as a function of the topics $\boldsymbol{\beta}_{1:K}$ and the multivariate Gaussian $(\boldsymbol{\mu}, \boldsymbol{\Sigma})$.

As in many latent variable models, we cannot compute the marginal likelihood of the data because of the latent structure that needs to be marginalized out. To address this issue, we use variational expectation–maximization (EM). In the E-step of traditional EM, one computes the posterior distribution of the latent variables given the data and current model parameters. In variational EM, we use the variational approximation to the posterior described in the previous section. Note that this is akin to Monte Carlo EM, where the E-step is approximated by a Monte Carlo approximation to the posterior [30].

Specifically, the objective function of variational EM is the likelihood bound given by summing equation (3) over the document collection $\{\boldsymbol{w}_1, \ldots, \boldsymbol{w}_D\}$,

$$\mathcal{L}(\boldsymbol{\mu}, \boldsymbol{\Sigma}, \boldsymbol{\beta}_{1:K}; \boldsymbol{w}_{1:D}) \geq \sum_{d=1}^{D} \mathrm{E}_{q_d}[\log p(\boldsymbol{\eta}_d, \boldsymbol{z}_d, \boldsymbol{w}_d | \boldsymbol{\mu}, \boldsymbol{\Sigma}, \boldsymbol{\beta}_{1:K})] + \mathrm{H}(q_d).$$

The variational EM algorithm is coordinate ascent in this objective function. In the E-step, we maximize the bound with respect to the variational parameters by performing variational inference for each document. In the M-step, we maximize the bound with respect to the model parameters. This amounts to maximum likelihood estimation of the topics and multivariate Gaussian using expected sufficient statistics, where the expectation is taken with respect to the variational distributions computed in the E-step,

$$\begin{aligned}
\widehat{\boldsymbol{\beta}}_i &\propto \sum_d \phi_{d,i} \mathbf{n}_d, \\
\widehat{\boldsymbol{\mu}} &= \frac{1}{D} \sum_d \boldsymbol{\lambda}_d, \\
\widehat{\boldsymbol{\Sigma}} &= \frac{1}{D} \sum_d I\nu_d^2 + (\boldsymbol{\lambda}_d - \widehat{\boldsymbol{\mu}})(\boldsymbol{\lambda}_d - \widehat{\boldsymbol{\mu}})^T,
\end{aligned}$$

where $\mathbf{n}_d$ is the vector of word counts for document $d$.

The E-step and M-step are repeated until the bound on the likelihood converges. In the analysis reported below, we run variational inference until the relative change in the probability bound of equation (3) is less than 0.0001%, and run variational EM until the relative change in the likelihood bound is less than 0.001%.

3.3. *Topic graphs.* As seen below, the ability of the CTM to model the correlation between topics yields a better fit of a document collection than LDA. But the covariance of the logistic normal model for topic proportions can also be used to visualize the relationships among the topics. In particular, the covariance matrix can be used to form a topic graph, where the nodes represent individual topics, and neighboring nodes represent highly related topics. In such settings, it is useful to have a mechanism to control the sparsity of the graph.

Recall that the graph encoding the independence relations in a Gaussian graphical model is specified by the zero pattern in the inverse covariance matrix. More precisely, if $X \sim N(\boldsymbol{\mu}, \boldsymbol{\Sigma})$ is a $K$-dimensional multivariate Gaussian, and $\boldsymbol{S} = \boldsymbol{\Sigma}^{-1}$ denotes the inverse covariance matrix, then we form a graph $\mathcal{G}(\boldsymbol{\Sigma}) = (V, E)$ with vertices $V$ corresponding to the random variables $X_1, \ldots, X_K$ and edges $E$ satisfying $(s,t) \in E$ if and only if $S_{st} \neq 0$. If $\mathcal{N}(s) = \{t : (s,t) \in E\}$ denotes the set of neighbors of $s$ in the graph, then the independence relation $X_s \perp X_u | X_{\mathcal{N}(s)}$ holds for any node $u \notin \mathcal{N}(s)$ that is not a neighbor of $s$.

Recent work of Meinshausen and Bühlmann [20] shows how the lasso [28] can be adapted to give an asymptotically consistent estimator of the graph $\mathcal{G}(\boldsymbol{\Sigma})$. The strategy is to regress each variable $X_s$ onto all of the other variables, imposing an $\ell_1$ penalty on the parameters to encourage sparsity. The nonzero components then serve as an estimate of the neighbors of $s$ in the graph.

In more detail, let $\boldsymbol{\kappa}_s = (\kappa_{s1}, \ldots, \kappa_{sK}) \in \mathbf{R}^K$ be the parameters of the lasso fit obtained by regressing $X_s$ onto $(X_t)_{t \neq s}$, with the parameter $\kappa_{ss}$ serving as the unregularized intercept. The optimization problem is

$$\widehat{\boldsymbol{\kappa}}_s = \arg\min_{\kappa} \tfrac{1}{2} \|X_s - \boldsymbol{X}_{\backslash s} \boldsymbol{\kappa}_s\|_2^2 + \rho_n \|\boldsymbol{\kappa}_{\backslash s}\|_1, \tag{5}$$

where $\boldsymbol{X}_{\backslash s}$ denotes the set of variables with $X_s$ replaced by the vector of all 1's, and $\boldsymbol{\kappa}_{\backslash s}$ denotes the vector $\boldsymbol{\kappa}_s$ with component $\kappa_{ss}$ removed. The estimated set of neighbors is then

$$\widehat{\mathcal{N}}(s) = \{t : \widehat{\kappa}_{st} \neq 0\}. \tag{6}$$

Meinshausen and Bühlmann [20] show that $\mathbf{P}(\widehat{\mathcal{N}}(s) = \mathcal{N}(s)) \to 1$ as the sample size $n$ increases, for a suitable choice of the regularization parameter $\rho_n$ satisfying $n\rho_n^2 - \log(K) \to \infty$. Moreover, the convergence is exponentially fast, and as a consequence, if $K = O(n^d)$ grows only polynomially with sample size, the estimated graph is the true graph with probability approaching one.

To adapt the Meinshausen–Bühlmann technique to the CTM, recall that we estimate the covariance matrix $\boldsymbol{\Sigma}$ using variational EM, where in the M-step we maximize the variational lower bound with respect to approximation computed in the E-step. For a given document $d$, the variational approximation to the posterior of $\boldsymbol{\eta}$ is a normal with mean $\boldsymbol{\lambda}_d \in \mathbf{R}^K$. We treat the standardized mean vectors $\{\boldsymbol{\lambda}_d\}$ as data, and regress each component onto the others with an $\ell_1$ penalty. Two simple procedures can be used to then form the graph edge set, by taking the conjunction or disjunction of the local neighborhood estimates:

$$(s,t) \in E^{\text{AND}} \qquad \text{in case } t \in \widehat{\mathcal{N}}(s) \text{ and } s \in \widehat{\mathcal{N}}(t), \tag{7}$$

$$(s,t) \in E^{\text{OR}} \qquad \text{in case } t \in \widehat{\mathcal{N}}(s) \text{ or } s \in \widehat{\mathcal{N}}(t). \tag{8}$$

Figure 2 shows an example of a topic graph constructed using this method, with edges $E^{\text{AND}}$ formed by intersecting the neighborhood estimates. Varying the regularization parameter $\rho_n$ allows control over the sparsity of the graph; the graph becomes increasingly sparse as $\rho_n$ increases.

**4. Analyzing Science.** JSTOR is an on-line archive of scholarly journals that scans bound volumes dating back to the 1600s and runs optical character recognition algorithms on the scans. Thus, JSTOR stores and indexes hundreds of millions of pages of noisy text, all searchable through the Internet. This is an invaluable resource to scholars.

The JSTOR collection provides an opportunity for developing exploratory analysis and useful descriptive statistics of large volumes of text data. As they are, the articles are organized by journal, volume and number. But the users of JSTOR would benefit from a topical organization of articles from different journals and automatic recommendations of similar articles to those known to be of interest.

In some modern electronic scholarly archives, such as the ArXiv (http://www.arxiv.org/), contributors provide meta-data with their manuscripts that describe and categorize their work to aid in such a topical exploration of the collection. In many text data sets, however, meta-data is unavailable. Moreover, there may be underlying topics and connections between articles that the authors or curators have not determined. To these ends, we analyzed a large portion of JSTOR's corpus of articles from *Science* with the CTM.

4.1. *Qualitative analysis of Science.* In this section we illustrate the possible applications of the CTM to automatic corpus analysis and browsing. We estimated a 100-topic model on the *Science* articles from 1990 to 1999 using the variational EM algorithm of Section 3.2. (C code that implements this algorithm can be found at the first author's web-site and STATLIB.) The total vocabulary size in this collection is 375,144 terms. We trim the 356,195 terms that occurred fewer than 70 times as well as 296 stop words, that is, words like "the," "but" or "with," which do not convey meaning. This yields a corpus of 16,351 documents, 19,088 unique terms and a total of 5.7M words.

Using the technique described in Section 3.3, we constructed a sparse graph ($\rho = 0.1$) of the connections between the estimated latent topics. Part of this graph is illustrated in Figure 2. (For space, we manually removed topics that occurred very rarely and those that captured nontopical content such as front matter.) This graph provides a snapshot of ten years of *Science*, and reveals different substructures of themes in the collection. A user interested in the brain can restrict attention to articles that use the neuroscience topics; a user interested in genetics can restrict attention to those articles in the cluster of genetics topics.

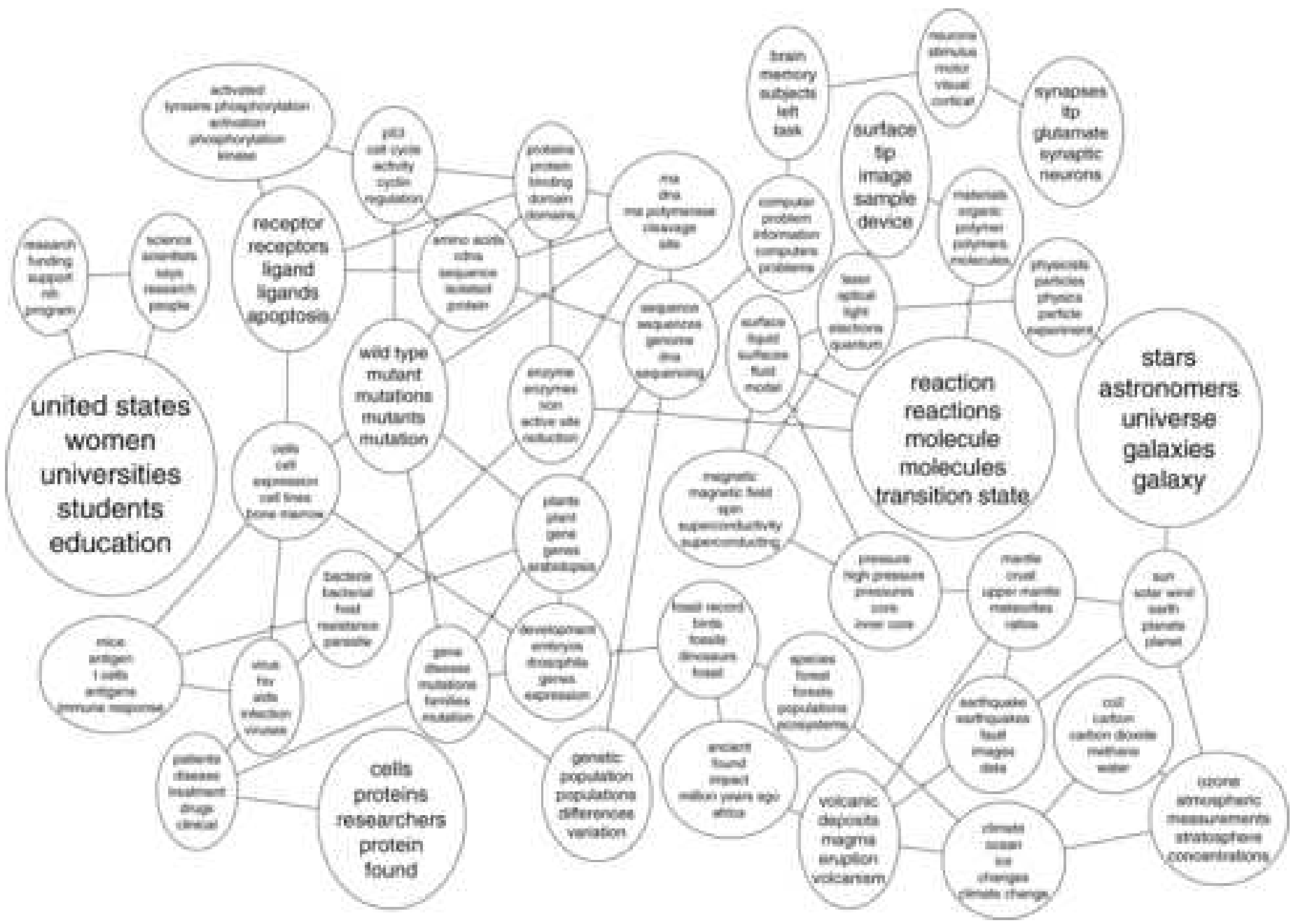

FIG. 2. *A portion of the topic graph learned from* 16,351 *OCR articles from Science (1990–1999). Each topic node is labeled with its five most probable phrases and has font proportional to its popularity in the corpus. (Phrases are found by permutation test.) The full model can be found in http://www.cs.cmu.edu/~lemur/science/ and on STATLIB.*

Further structure is revealed at the document level, where each document is associated with a latent vector of topic proportions. The posterior distribution of the proportions can be used to associate documents with latent topics. For example, the following are the top five articles associated with the topic whose most probable vocabulary items are "laser, optical, light, electron, quantum":

1. "Vacuum Squeezing of Solids: Macroscopic Quantum States Driven by Light Pulses" (1997).
2. "Superradiant Rayleigh Scattering from a Bose–Einstein Condensate" (1999).
3. "Physics and Device Applications of Optical Microcavities" (1992).
4. "Photon Number Squeezed States in Semiconductor Lasers" (1992).
5. "A Well-Collimated Quasi-Continuous Atom Laser" (1999).

Moreover, we can use the expected distance between per-document topic proportions to identify other documents that have similar topical content. We use the expected Hellinger distance, which is a symmetric distance between distributions. Consider two documents $i$ and $j$,

$$\begin{aligned}\mathrm{E}[d(\boldsymbol{\theta}_i, \boldsymbol{\theta}_j)] &= \mathrm{E}_q\left[\sum_k (\sqrt{\theta_{ik}} - \sqrt{\theta_{jk}})^2\right] \\ &= 2 - 2\sum_k \mathrm{E}_q[\sqrt{\theta_{ik}}]\mathrm{E}_q\left[\frac{\theta_{jk}}{\sqrt{\theta_{jk}}}\right],\end{aligned}$$

where all expectations are taken with respect to the variational posterior distributions (see Section 3.1). One example of this application of the latent variable analysis is illustrated in Figure 3.

The interested reader is invited to visit http://www.cs.cmu.edu/~lemur/science/ to interactively explore this model, including the topics, their connections, the articles that exhibit them and the expected Hellinger similarity between articles.

4.2. *Quantitative comparison to latent Dirichlet allocation.* We compared the logistic normal to the Dirichlet by fitting a smaller collection of articles to CTM and LDA models of varying numbers of topics. This collection contains the 1,452 documents from 1960; we used a vocabulary of 5,612 words after pruning common function words and terms that occur once in the collection. Using ten-fold cross validation, we computed the log probability of the held-out data given a model estimated from the remaining data. A better model of the document collection will assign higher probability to the held out data. To avoid comparing bounds, we used importance sampling to compute the log probability of a document where the fitted variational distribution is the proposal.

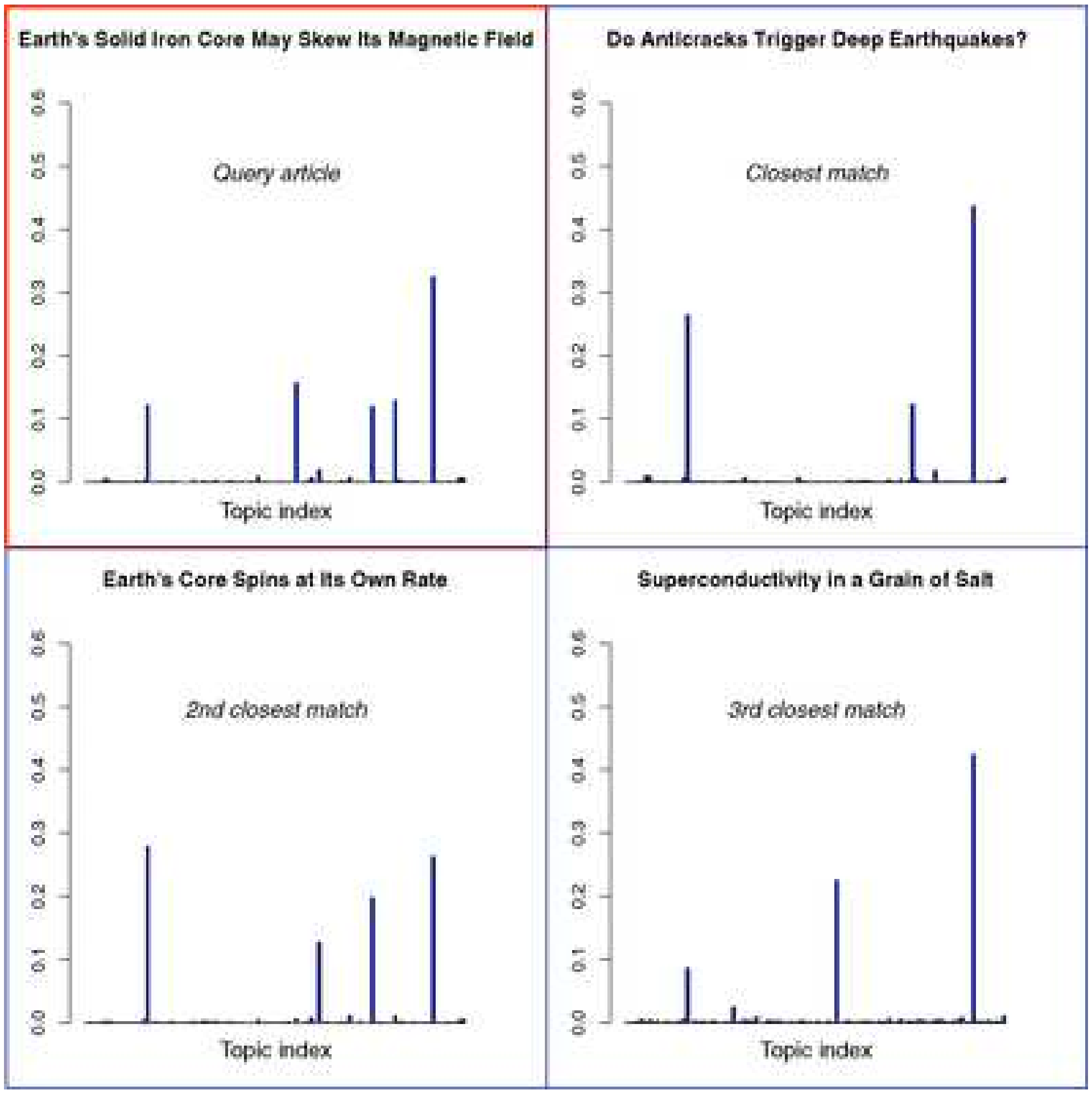

Fig. 3. *Using the Hellinger distance to find similar articles to the query article "Earth's Solid Iron Core May Skew Its Magnetic Field." Illustrated are the top three articles by Hellinger distance to the query article and the expected posterior topic proportions for each article. Notice that each document somehow combines geology and physics.*

Figure 4 illustrates the average held out log probability for each model and the average difference between them. The CTM provides a better fit than LDA and supports more topics; the likelihood for LDA peaks near 30 topics, while the likelihood for the CTM peaks close to 90 topics. The means and standard errors of the *difference* in log-likelihood of the models is shown at right; this indicates that the CTM always gives a better fit.

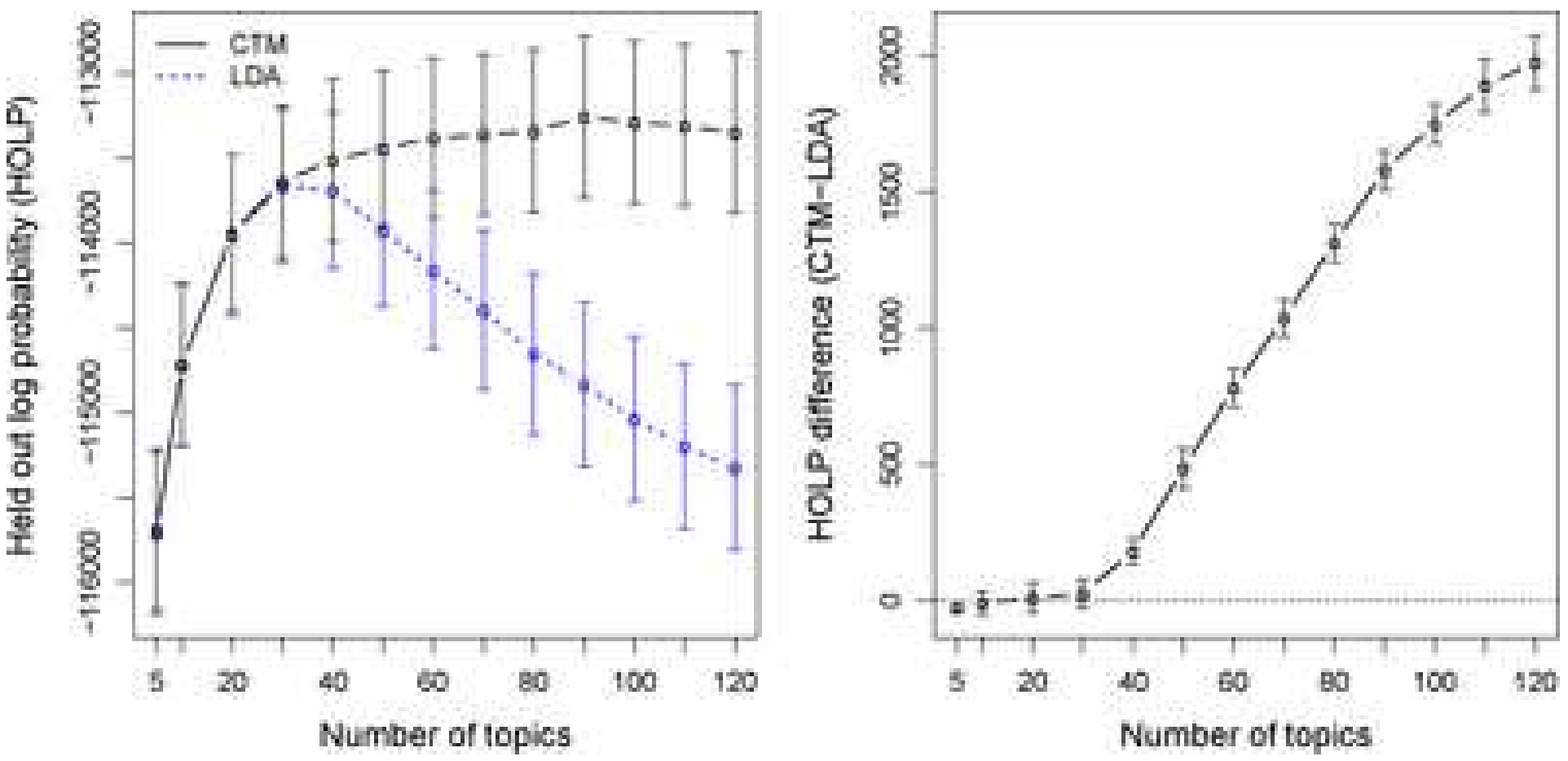

FIG. 4. (Left) *The 10-fold cross-validated held-out log probability of the 1960 Science corpus, computed by importance sampling. The CTM supports more topics than LDA. See figure at right for the standard error of the difference.* (Right) *The mean difference in held-out log probability. Numbers greater than zero indicate a better fit by the CTM.*

Another quantitative evaluation of the relative strengths of LDA and the CTM is how well the models predict the remaining words of a document after observing a portion of it. Specifically, we observe $P$ words from a document and are interested in which model provides a better predictive distribution of the remaining words $p(w|w_{1:P})$. To compare these distributions, we use *perplexity*, which can be thought of as the effective number of equally likely words according to the model. Mathematically, the perplexity of a word distribution is defined as the inverse of the per-word geometric average of the probability of the observations,

$$\mathrm{Perp}(\Phi) = \left(\prod_{d=1}^{D} \prod_{i=P+1}^{N_d} p(w_i|\Phi, w_{1:P})\right)^{-1/(\sum_{d=1}^{D}(N_d-P))},$$

where $\Phi$ denotes the model parameters of an LDA or CTM model. Note that lower numbers denote more predictive power.

The plot in Figure 5 compares the predictive perplexity under LDA and the CTM for different numbers of words randomly observed from the documents. When a small number of words have been observed, there is less uncertainty about the remaining words under the CTM than under LDA—the perplexity is reduced by nearly 200 words, or roughly 10%. The reason is that after seeing a few words in one topic, the CTM uses topic correlation to infer that words in a related topic may also be probable. In contrast, LDA cannot predict the remaining words as well until a large portion of the document has been observed so that all of its topics are represented.

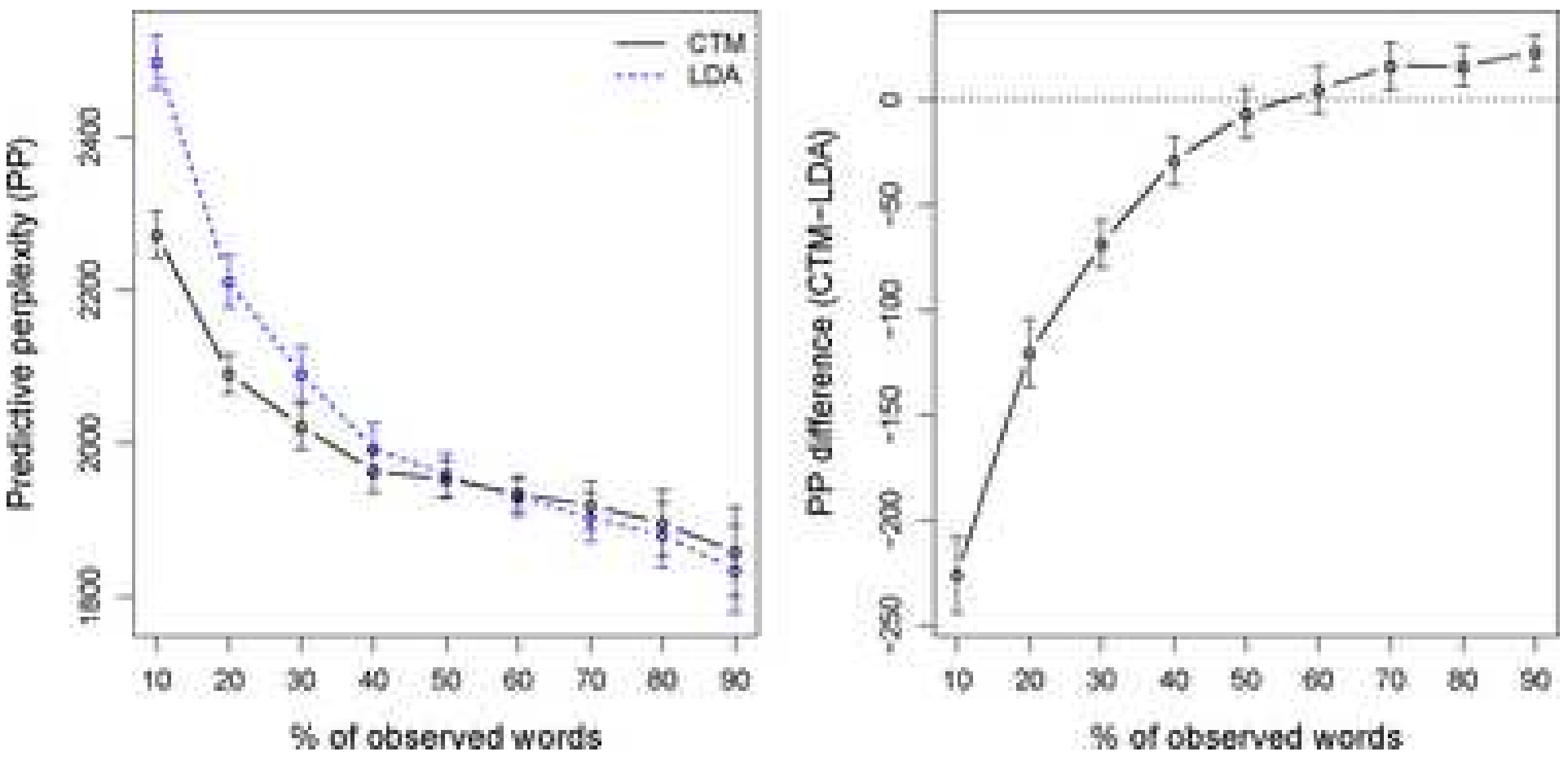

FIG. 5. (Left) *The* 10-*fold cross-validated predictive perplexity for partially observed held-out documents from the 1960 Science corpus* ($K = 50$). *Lower numbers indicate more predictive power from the CTM.* (Right) *The mean difference in predictive perplexity. Numbers less than zero indicate better prediction from the CTM.*

**5. Summary.** We have developed a hierarchical topic model of documents that replaces the Dirichlet distribution of per-document topic proportions with a logistic normal. This allows the model to capture correlations between the occurrence of latent topics. The resulting correlated topic model gives better predictive performance and uncovers interesting descriptive statistics for facilitating browsing and search. Use of the logistic normal, while more complex, may have benefit in the many applications of Dirichlet-based mixed membership models.

One issue that we did not thoroughly explore is model selection, that is, choosing the number of topics for a collection. In other topic models, nonparametric Bayesian methods based on the Dirichlet process are a natural suite of tools because they can accommodate new topics as more documents are observed. (The nonparametric Bayesian version of LDA is exactly the hierarchical Dirichlet process [27].) The logistic normal, however, does not immediately give way to such extensions. Tackling the model selection issue in this setting is an important area of future research.

## APPENDIX: DETAILS OF VARIATIONAL INFERENCE

*Variational objective.* Before deriving the optimization procedure, we put the objective function equation (3) in terms of the variational parameters. The first term is

$$\begin{aligned} &\mathrm{E}_q[\log p(\boldsymbol{\eta}|\boldsymbol{\mu}, \boldsymbol{\Sigma})] \\ &\quad = \frac{1}{2}\log|\boldsymbol{\Sigma}^{-1}| - \frac{K}{2}\log 2\pi - \frac{1}{2}\mathrm{E}_q[(\boldsymbol{\eta}-\boldsymbol{\mu})^T\boldsymbol{\Sigma}^{-1}(\boldsymbol{\eta}-\boldsymbol{\mu})], \end{aligned} \tag{9}$$

where

$$\begin{aligned} &\mathrm{E}_q[(\boldsymbol{\eta}-\boldsymbol{\mu})^T\boldsymbol{\Sigma}^{-1}(\boldsymbol{\eta}-\boldsymbol{\mu})] \\ &\quad = \mathrm{Tr}(\mathrm{diag}(\nu^2)\boldsymbol{\Sigma}^{-1}) + (\boldsymbol{\lambda}-\boldsymbol{\mu})^T\boldsymbol{\Sigma}^{-1}(\boldsymbol{\lambda}-\boldsymbol{\mu}). \end{aligned} \tag{10}$$

The nonconjugacy of the logistic normal to multinomial leads to difficulty in computing the second term of equation (3), the expected log probability of a topic assignment

$$\mathrm{E}_q[\log p(z_n|\boldsymbol{\eta})] = \mathrm{E}_q[\boldsymbol{\eta}^T z_n] - \mathrm{E}_q\left[\log\left(\sum_{i=1}^K \exp\{\eta_i\}\right)\right]. \tag{11}$$

To preserve the lower bound on the log probability, we upper bound the negative log normalizer with a Taylor expansion:

$$\mathrm{E}_q\left[\log\left(\sum_{i=1}^K \exp\{\eta_i\}\right)\right] \le \zeta^{-1}\left(\sum_{i=1}^K \mathrm{E}_q[\exp\{\eta_i\}]\right) - 1 + \log(\zeta), \tag{12}$$

where we have introduced a new variational parameter $\zeta$. The expectation $\mathrm{E}_q[\exp\{\eta_i\}]$ is the mean of a log normal distribution with mean and variance obtained from the variational parameters $\{\lambda_i, \nu_i^2\}$: $\mathrm{E}_q[\exp\{\eta_i\}] = \exp\{\lambda_i + \nu_i^2/2\}$ for $i \in \{1, \dots, K\}$. This is a simpler approach than the more flexible, but more computationally intensive, method taken in [25].

Using this additional bound, the second term of equation (3) is

$$\mathrm{E}_q[\log p(z_n|\boldsymbol{\eta})] = \sum_{i=1}^K \lambda_i\phi_{n,i} - \zeta^{-1}\left(\sum_{i=1}^K \exp\{\lambda_i + \nu_i^2/2\}\right) + 1 - \log\zeta. \tag{13}$$

The third term of equation (3) is

$$\mathrm{E}_q[\log p(w_n|z_n, \boldsymbol{\beta})] = \sum_{i=1}^K \phi_{n,i}\log\beta_{i,w_n}. \tag{14}$$

The fourth term is the entropy of the variational distribution:

$$\sum_{i=1}^K \tfrac{1}{2}(\log\nu_i^2 + \log 2\pi + 1) - \sum_{n=1}^N\sum_{i=1}^k \phi_{n,i}\log\phi_{n,i}. \tag{15}$$

Note that the additional variational parameter $\zeta$ is not needed to compute this entropy.

*Coordinate ascent optimization.* Finally, we maximize the bound in equation (3) with respect to the variational parameters $\lambda_{1:K}$, $\nu_{1:K}$, $\phi_{1:N}$ and $\zeta$. We use a coordinate ascent algorithm, iteratively maximizing the bound with respect to each parameter.

First, we maximize equation (3) with respect to $\zeta$, using the second bound in equation (12). The derivative with respect to $\zeta$ is

$$f'(\zeta) = N\left(\zeta^{-2}\left(\sum_{i=1}^{K}\exp\{\lambda_i + \nu_i^2/2\}\right) - \zeta^{-1}\right), \tag{16}$$

which has a maximum at

$$\widehat{\zeta} = \sum_{i=1}^{K}\exp\{\lambda_i + \nu_i^2/2\}. \tag{17}$$

Second, we maximize with respect to $\boldsymbol{\phi}_n$. This yields a maximum at

$$\widehat{\phi}_{n,i} \propto \exp\{\lambda_i\}\beta_{i,w_n}, \qquad i \in \{1,\ldots,K\}, \tag{18}$$

which is an application of variational inference updates within the exponential family [5, 7, 31].

Third, we maximize with respect to $\lambda_i$. Equation (3) is not amenable to analytic maximization. We use the conjugate gradient algorithm with derivative

$$dL/d\boldsymbol{\lambda} = -\boldsymbol{\Sigma}^{-1}(\boldsymbol{\lambda} - \boldsymbol{\mu}) + \sum_{n=1}^{N}\phi_{n,1:K} - (N/\zeta)\exp\{\boldsymbol{\lambda} + \boldsymbol{\nu}^2/2\}. \tag{19}$$

Finally, we maximize with respect to $\nu_i^2$. Again, there is no analytic solution. We use Newton's method for each coordinate with the constraint that $\nu_i > 0$,

$$dL/d\nu_i^2 = -\boldsymbol{\Sigma}_{ii}^{-1}/2 - N/2\zeta\exp\{\lambda_i + \nu_i^2/2\} + 1/(2\nu_i^2). \tag{20}$$

Iterating between the optimizations of $\boldsymbol{\nu}$, $\boldsymbol{\lambda}$, $\boldsymbol{\phi}$ and $\zeta$ defines a coordinate ascent algorithm on equation (3). (In practice, we optimize with respect to $\zeta$ in between optimizations for $\boldsymbol{\nu}$, $\boldsymbol{\lambda}$ and $\boldsymbol{\phi}$.) Though each coordinate's optimization is convex, the variational objective is not convex with respect to the ensemble of variational parameters. We are only guaranteed to find a local maximum, but note that this is still a bound on the log probability of a document.

**Acknowledgments.** We thank two anonymous reviewers for their excellent suggestions for improving the paper. We thank JSTOR for providing access to journal material in the JSTOR archive. We thank Jon McAuliffe and Nathan Srebro for useful discussions and comments. A preliminary version of this work appears in [9].

Computer Science Department
Princeton University
Princeton, New Jersey 08540
USA
E-mail: blei@cs.princeton.edu

Computer Science Department
Machine Learning Department
Carnegie Mellon University
Pittsburgh, Pennsylvania 15213
USA
E-mail: lafferty@cs.cmu.edu